\documentclass[journal]{IEEEtran}

\usepackage{blindtext}
\usepackage{graphicx}

\usepackage{bm}
\usepackage{amsfonts}
\usepackage{cite}

\ifCLASSINFOpdf
\else
\fi
\usepackage{color}

\usepackage{amsmath}
 \usepackage[caption=false,font=footnotesize]{subfig}
\begin{document}
%
\title{Robust MIMO Beamforming for Cellular and Radar Coexistence}
%
%
%

\author{Fan Liu,~\IEEEmembership{Student~Member,~IEEE,}
        Christos Masouros,~\IEEEmembership{Senior~Member,~IEEE,}
        Ang Li,~\IEEEmembership{Student~Member,~IEEE,}
        and~Tharmalingam Ratnarajah,~\IEEEmembership{Senior~Member,~IEEE}
\thanks{Manuscript received January 25, 2017; revised March 13, 2017; accepted April 9, 2017. This work was supported by the Engineering and Physical Sciences Research Council (EPSRC) project EP/M014150/1 and the China Scholarship Council (CSC). The editor coordinating the review of this letter and approving it for publication was Prof. M. Nafie.}
\thanks{Fan Liu is with the School of Information and Electronics, Beijing Institute of Technology, Beijing, 100081, China (e-mail: liufan92@bit.edu.cn).
Fan Liu, Christos Masouros and Ang Li are with the Department of Electronic and Electrical Engineering, University College London, London, WC1E 7JE, UK (e-mail: chris.masouros@ieee.org, ang.li.14@ucl.ac.uk). 
Tharmalingam Ratnarajah is with the Institute for Digital Communications, School of Engineering, The University of Edinburgh, Edinburgh, EH9 3JL, UK (e-mail: t.ratnarajah@ed.ac.uk).
}

}

\maketitle
\begin{abstract}
 In this letter, we consider the coexistence and spectrum sharing between downlink multi-user multiple-input-multiple-output (MU-MIMO) communication and a MIMO radar. For a given performance requirement of the downlink communication system, we design the transmit beamforming such that the detection probability of the radar is maximized. While the original optimization problem is non-convex, we exploit the monotonically increasing relationship of the detection probability with the non-centrality parameter of the resulting probability distribution to obtain a convex lower-bound optimization. The proposed beamformer is designed to be robust to imperfect channel state information (CSI). Simulation results verify that the proposed approach facilitates the coexistence between radar and communication links, and illustrates a scalable trade-off between the two systems' performance.
\end{abstract}



\section{Introduction}
%
%
%
%
\IEEEPARstart{T}{o} address the explosive growth of wireless communication devices and services, a broadband plan has been agreed to free additional spectrum that is currently exclusive for military and governmental operations\cite{federal2010connecting}. Typically, this spectrum is occupied by air surveillance and weather radar systems, and henceforth spectrum sharing between radar and communication has drawn much attention as an enabling solution \cite{7782415}. While policy and regulations my delay the practical application of such solutions, research efforts are well under way to address the practical implementation of radar and communication coexistence. In \cite{6331681}, Opportunistic Spectrum Sharing (OSS) between cellular system and rotating radar has been considered, where the communication system are allowed to transmit signals when the space and frequency spectra are not occupied by the radar. Although the OSS method is straightforward, it does not allow radar and communication to work simultaneously. Additionally, traditional rotating radar will soon be replaced by MIMO radar in the near future due to the advantages of waveform diversity and higher detection capability \cite{4350230}. In recent years, several methods that consider the coexistence between MIMO radar and MIMO communication have been proposed, among which the Null Space Projection (NSP) method has been widely discussed \cite{6831613,7089157}. More relevant to this work, optimization techniques have also been proposed to solve the problem. In \cite{7485158}, authors optimize the Signal-to-Interference-plus-Noise-Ratio (SINR) of radar subject to power and capacity constraints. Related work discusses the coexistence between MIMO-Matrix Completion (MIMO-MC) radar and MIMO communication system, where the radar beamforming matrix and communication covariance matrix are jointly optimized\cite{7470514}. Similar work has been done in \cite{7511108}, where the transmit beamforming design for the base station (BS) based on Linearly Constrained Minimum Variance (LCMV) optimization is proposed. Nevertheless, all of these works assume that CSI is perfectly known by radar or BS, which is not possible in practical scenarios. While robust beamformers exist in the broader area of cognitive radio networks for unicast and multicast transmission\cite{4787135,6373750}, robust radar-specific coexistence solutions are yet to be explored in the related literature.
\\\indent In this letter, we consider the transmit beamforming for spectrum sharing between downlink MU-MIMO communication and colocated MIMO radar. Focusing on a radar-specific optimization, we maximize the detection probability of radar while guaranteeing the transmit power budget of the BS and the received SINR of each downlink user. The beamforming design is initially formulated as an optimization problem under the perfect CSI assumption. Since the objective function is non-concave, we then optimize its lower bound instead. We further consider two optimization approaches where both the communication channel and interference channel are subject to CSI quantization errors. The proposed problems can be transformed into Semidefinite Programs (SDP) and solved by Semidefinite Relaxation (SDR) techniques. Simulation results validate the effectiveness of the proposed beamforming approach under the coexistence scenario for both perfect and imperfect CSI cases.

\section{System Model}
 
We consider a Time Division Duplex (TDD) downlink MU-MIMO communication system that coexists with a MIMO radar on the same frequency band. As shown in Fig. 1, an \emph{N}-antenna BS transmits signals to \emph{K} single-antenna users. Meanwhile, a MIMO radar with $ M_{t} $ TX and $ M_{r} $ RX antennas is detecting a point-like target in the far-field. The received signal at the \emph{i}-th user is 
\begin{equation}
    {{y}^C_i[l]} = {\bf{h}}_i^T\sum\limits_{k = 1}^K {{{\bf{t}}_k}{d}_k[l]}  + \sqrt{P_R}{\bf{f}}_i^T{{\bf s}_l} + {n}_i[l], i = 1,2,...,K,
\end{equation}
where $ {\bf{h}}_{i} \in {\mathbb{C}^{\emph{N} \times 1}} $, $ {\bf{f}}_{i} \in {\mathbb{C}^{{{\emph{M}}_{t}} \times 1}} $, $ {\bf{t}}_{i} \in {\mathbb{C}^{\emph{N} \times 1}} $, $ {{d}}_i[l] $ and $ {{n}}_i[l] \sim {\mathcal{C}}{\mathcal{N}}\left( {0,{\sigma _C^2}} \right)$ denote the communication channel vector, the interference channel vector from radar, the beamforming vector, the communication symbol and the received noise for the  \emph{i}-th user respectively.  $ l = 1,2,...,L $ is the symbol duration index, and $ L $ is the length of the communication frame. $ {\bf{S}} = \left[ {\bf{s}}_1,{\bf{s}}_2,...,{\bf{s}}_L \right] \in {\mathbb{C}^{M_t \times L}} $ denotes the radar transmit waveforms, with ${\bf{s}}_l$ being the \emph{l}-th snapshot across the transmit antennas. $ P_R $ is the power of the radar signals. Without loss of generality, we assume that the communication symbol has unit power, i.e., $ \mathbb{E}\left[ {{{\left| {{{{d}}_k[l]}} \right|}^2}} \right] = 1 $, where $ \mathbb{E} $ denotes the ensemble average. It is also assumed that MIMO radar uses orthogonal waveforms, i.e., $\mathbb{E}\left[ {{{\bf s}_l}{{\bf s}^H_l}} \right] = \frac{1}{L}\sum\limits_{l = 1}^L {{{\bf s}_l}{{\bf s}^H_l}}  = {\mathbf{I}}$. The received SINR at the \emph{i}-th user is thus given as
\begin{equation}
    {\gamma _i} = \frac{{{{\left| {{\bf{h}}_i^T{{\bf{t}}_i}} \right|}^2}}}{{\sum\limits_{k = 1,k \ne i}^K {{{\left| {{\bf{h}}_i^T{{\bf{t}}_k}} \right|}^2} + {P_R}{{\left\| {{{\bf{f}}_i}} \right\|}^2} + \sigma _C^2} }}, \forall i.
\end{equation}
\\\indent Considering the echo wave in a single range-Doppler bin of the radar detector, at the \emph{l}-th snapshot, the discrete signal vector ${{\bf{y}}^{R}_l}$ received by radar is given as
\begin{equation}
    {{\bf{y}}^R_l}=\alpha \sqrt{P_R} {\bf{A}}\left( \theta  \right){{\bf s}_l} + {{\bf{G}}^T}\sum\limits_{k = 1}^K {{{\bf{t}}_k}{d_k}\left[ l \right]} + {\bf{w}}_l,
\end{equation}
where $ {\bf G }=\left[{\bf{g}}_1,{\bf{g}}_2,...,{\bf{g}}_{M_r} \right] \in {\mathbb{C}}^{N\times M_r}$ is the interference channel matrix between BS and radar RX, $\theta$ is the azimuth angle of the target, $\alpha$ is the complex path loss of the radar-target-radar path, $ {\bf{w}}_l=\left[w_1\left[l\right], w_2\left[l\right],..., w_{M_r}\left[l\right]\right]^T \in {\mathbb{C}}^{M_r\times 1} $ is the received noise vector at the \emph{l}-th snapshot with $w_m[l]\sim {\mathcal{C}}{\mathcal{N}}\left( {0,{\sigma _R^2}} \right), \forall m$, $ {\bf{A}}\left( \theta  \right) = {{\bf{a}}_R}\left( \theta  \right){\bf{a}}_T^T\left( \theta  \right) $, in which $ {{\bf{a}}_T}\left( \theta  \right) \in {\mathbb{C}^{{M_t} \times 1}} $ and $ {{\bf{a}}_R}\left( \theta  \right) \in {\mathbb{C}^{{M_r} \times 1}}  $ are transmit and receive steering vectors of radar antenna array. In this letter, the model in\cite{1703855} is used, for which
\begin{equation}
\begin{gathered}
  {M_r} = {M_t} = M, 
  \;\;{{\mathbf{a}}_R}\left( \theta  \right) = {{\mathbf{a}}_T}\left( \theta  \right) = {\mathbf{a}}\left( \theta  \right), \hfill \\
  {{\mathbf{A}}_{im}}\left( \theta  \right) = {{\mathbf{a}}_i}\left( \theta  \right){{\mathbf{a}}_m}\left( \theta  \right) = \exp \left( { - j\omega {\tau _{im}}\left( \theta  \right)} \right) \hfill \\
   \;\;\;\;\;\;\;\;\;\;\;\;\; = \exp \left( { - j\frac{{2\pi }}{\lambda }{{\left[ {\sin \left( \theta  \right);\cos \left( \theta  \right)} \right]}^T}\left( {{{\mathbf{x}}_i} + {{\mathbf{x}}_m}} \right)} \right), \hfill \\ 
\end{gathered} 
\end{equation}
where $ \omega $ and $ \lambda $ denote the frequency and the wavelength of the carrier, $ {{\bf A}_{im}}\left( \theta  \right) $ is the $ i $-th element at the $ m $-th column of the matrix $\bf A$, which is the total phase delay of the signal, transmitted by the \emph{i}-th element and received by the \emph{m}-th element of the antenna array, $ {{\bf{x}}_i} = \left[ {x_i^1;x_i^2} \right] $ is the location of the \emph{i}-th element of the antenna array. In the above model, we assume that $ {\bf{H}}=\left[{\bf{h}}_1, {\bf{h}}_2,...,{\bf{h}}_K \right] $, $ {\bf{F}}=\left[{\bf{f}}_1, {\bf{f}}_2,...,{\bf{f}}_K \right] $ and $\bf G$ are flat Rayleigh fading and independent with each other and can be estimated by the BS through the pilot symbols. Note that for a typical TDD downlink, users will remain silence when BS is transmitting signals, so the radar only receives interference from the BS. For convenience, the index $ l $ is omitted in the rest of the letter.
\\\indent The interference from BS to radar RX will affect the detection probability of radar. Under the Neyman-Pearson criterion, by using the Generalized Likelihood Ratio Test (GLRT), the asymptotic radar detection probability $P_D$ is given as \footnote{It should be highlighted that the derivation in \cite{1703855} is for the scenario with white Gaussian noise only while the proposed model in (3) includes both interference and noise. However, it can be shown that the resultant interference-plus-noise is still i.i.d. Gaussian distributed, but with a non-identity covariance matrix. We therefore apply a whitening-filter to normalize the interference-plus-noise, such that the derivation in \cite{1703855} is still valid.}\cite{1703855}
\begin{equation}
{P_D} = 1 - {\mathfrak{F}_{\mathcal{X}_2^2\left( \rho  \right)}}\left( {\mathfrak{F}_{\mathcal{X}_2^2}^{ - 1}\left( {1 - {P_{FA}}} \right)} \right),
\end{equation}
where $ P_{FA} $ is radar's probability of false alarm, $ {\mathfrak{F}_{\mathcal{X}_2^2\left( \rho  \right)}} $ is the non-central chi-square Cumulative Distribution Function (CDF) with 2 Degrees of Freedom (DoF), $ {\mathfrak{F}_{\mathcal{X}_2^2}^{- 1}} $ is the inverse function of chi-square CDF with 2 DoFs. Let $ {\mathbf{\tilde T}} = \sum\limits_{k = 1}^K {{{\mathbf{t}}_k}{\mathbf{t}}_k^H}  $, the non-centrality parameter $ \rho $ for $ {\mathcal{X}_2^2\left( \rho  \right)} $ is given by \cite{kay1998fundamentals}
\begin{equation}
\rho  = {|\alpha| ^2}L{P_R}\text{tr} \left( {{\mathbf{A}}{{\mathbf{A}}^H}{{\left( {{{\mathbf{G}}^T}{\mathbf{\tilde T}}{{\mathbf{G}}^*} + \sigma _R^2{\mathbf{I}}} \right)}^{ - 1}}} \right) .
\end{equation}

\begin{figure}
    \centering
    \includegraphics[width=0.55\columnwidth]{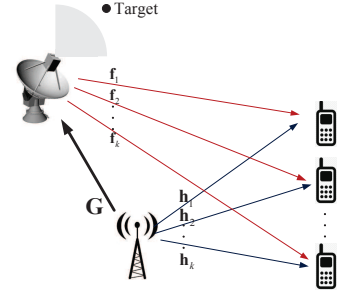}
    \caption{Spectrum sharing scenario.}
    \label{fig:1}
\end{figure}

\section{Proposed Beamforming Optimization}
The average transmit power per frame of the BS is 
\begin{equation}
    {P_C} = \sum\limits_{k = 1}^K {{{\left\| {{{\bf{t}}_k}} \right\|}^2}}=\sum\limits_{k = 1}^K {{\text{tr}}\left( {{{\mathbf{t}}_k}{{\mathbf{t}}_k^H}} \right)}=\sum\limits_{k = 1}^K {{\text{tr}}\left( {{{\mathbf{T}}_k}} \right)},
\end{equation}
where ${\mathbf{T}}_k={{{\mathbf{t}}_k}{{\mathbf{t}}_k^H}}$. The goal is to maximize the detection performance of radar while guaranteeing the received SINR per user and the power budget for the BS. We first consider the optimization with perfect CSI, followed by two optimization approaches, the upper bound minimization and the weighted minimization with norm-bounded CSI errors.
\subsection{Beamforming for Perfect CSI}
The optimization problem can be formulated as
\begin{equation}
\begin{array}{*{20}{l}}
  {{\cal{P}}_0}:\mathop {\max }\limits_{{{\mathbf{t}}_k}} \;\;{P_D} \hfill \\
  s.t.\;\;{\gamma _i} \ge {\Gamma _i},\forall i, \hfill \\
  \;\;\;\;\;\;\;{P_C} \le {P_0}, \hfill \\ 
\end{array}
\end{equation}
where $ \Gamma_i $ is the required SINR of the \emph{i}-th communication user, $ P_0 $ is the power budget of the BS, $ \gamma_i$, $P_D$ and $P_C$ are defined as (2), (5) and (7) respectively. It is well-known that $P_D$ is a monotonically increasing function with respect to $\rho$\cite{kay1998fundamentals}, thus problem ${{\cal{P}}_0}$ can be equivalently formulated as
\begin{equation}
\begin{array}{*{20}{l}}
  {{\cal{P}}_1}:\mathop {\max }\limits_{{{\mathbf{t}}_k}} \;\;\text{tr} \left( {{\mathbf{A}}{{\mathbf{A}}^H}{{\left( {{{\mathbf{G}}^T}{\mathbf{\tilde T}}{{\mathbf{G}}^*} + \sigma _R^2{\mathbf{I}}} \right)}^{ - 1}}} \right) \hfill \\
  s.t.\;\;{\gamma _i} \ge {\Gamma _i},\forall i, \hfill \\
  \;\;\;\;\;\;\;{P_C} \le {P_0}. \hfill \\ 
\end{array}
\end{equation}
As the objective function is non-concave, we consider a relaxation of optimizing its lower bound. Let ${\bf J}={{{\mathbf{G}}^T}{\mathbf{\tilde T}}{{\mathbf{G}}^*} + \sigma _R^2{\mathbf{I}}}$. Noting that both $\bf J$ and ${\mathbf{A}}{{\mathbf{A}}^H}$ are positive-definite, we have
\begin{equation}
\text{tr} \left( {{\mathbf{A}}{{\mathbf{A}}^H}{{\mathbf{J}}^{ - 1}}{\mathbf{J}}} \right) \le \text{tr} \left( {{\mathbf{A}}{{\mathbf{A}}^H}{{\mathbf{J}}^{ - 1}}} \right)\text{tr} \left( {\mathbf{J}} \right)
\end{equation}
\begin{equation}
\begin{array}{*{20}{l}}
\displaystyle \Rightarrow\text{tr} \left( {{\mathbf{A}}{{\mathbf{A}}^H}{{\mathbf{J}}^{ - 1}}} \right) \ge \frac{{\text{tr} \left( {{\mathbf{A}}{{\mathbf{A}}^H}} \right)}}{{\text{tr} \left( {\mathbf{J}} \right)}}  = \frac{{{M^2}}}{{\text{tr} \left( {{{\mathbf{G}}^T}{\mathbf{\tilde T}}{{\mathbf{G}^*}}} \right) + M\sigma _R^2}}.
\end{array}
\end{equation}
Based on (11), ${{\cal{P}}_1}$ can be relaxed as
\begin{equation}
\begin{array}{*{20}{l}}
  {{\cal{P}}_2}:\mathop {\min }\limits_{{{\mathbf{t}}_k}} \;\;\text{tr} \left( {{{\mathbf{G}}^T}{\mathbf{\tilde T}}{{\mathbf{G}}^*}} \right) \hfill \\
  s.t.\;\;{\gamma _i} \ge {\Gamma _i},\forall i, \hfill \\
  \;\;\;\;\;\;\;{P_C} \le {P_0}. \hfill \\ 
\end{array}
\end{equation}
${{\cal{P}}_2}$ is non-convex, and is equivalent to minimizing the total interference power from BS to radar. Fortunately, it can be efficiently solved by SDR technique. We refer readers to \cite{5447068} for details on this topic.
\subsection{Upper Bound Minimization for Imperfect CSI}
Let us first model the channel vectors as
\begin{equation}
    \begin{gathered}
  {{\mathbf{h}}_i} = {{{\mathbf{\hat h}}}_i} + {{\mathbf{e}}_{hi}}, {{\mathbf{f}}_i} = {{{\mathbf{\hat f}}}_i} + {{\mathbf{e}}_{fi}},\forall i, \hfill \\
  {{\mathbf{g}}_m} = {{{\mathbf{\hat g}}}_m} + {{\mathbf{e}}_{gm}},\forall m, \hfill \\
\end{gathered}
\end{equation}
where $ {\bf{\hat h}}_i $, $ {\bf{\hat g}}_m $ and $ {\bf{\hat f}}_i $ denote the estimated channel vectors known to the BS, $ {\bf{e}}_{hi} $, $ {\bf{e}}_{gm} $ and $ {\bf{e}}_{fi} $ denote the CSI uncertainty within the spherical sets $ {{\cal{U}}_{hi}}=\left\{ {{{\bf{e}}_{hi}}|{{\left\| {{{\bf{e}}_{hi}}} \right\|}^2} \le \delta _{hi}^2} \right\}  $, $ {{\cal{U}}_{gm}}=\left\{ {{{\bf{e}}_{gm}}|{{\left\| {{{\bf{e}}_{gm}}} \right\|}^2} \le \delta _{gm}^2} \right\}  $ and $ {{\cal{U}}_{fi}}=\left\{ {{{\bf{e}}_{fi}}|{{\left\| {{{\bf{e}}_{fi}}} \right\|}^2} \le \delta _{fi}^2} \right\}  $. This model is reasonable for scenarios where the CSI is quantized at the receiver and fed back to the BS. Particularly, if the quantizer is uniform, the quantization error region can be covered by spheres of given sizes \cite{4586299}.
\\\indent It is assumed that BS has no knowledge about the error vectors except for the bounds of their norms. Given the partially known $\bf G$, following the process of \cite{4787135}, the upper bound of the interference power for the \emph{m}-th radar antenna is given by
\begin{equation}
\begin{gathered}
  {\left| {{\mathbf{g}}_m^T\sum\limits_{k = 1}^K {{{\mathbf{t}}_k}}{d_k} } \right|^2} 
   \le \sum\limits_{k = 1}^K {\text{tr} \left( {{\mathbf{\hat g}}_m^*{\mathbf{\hat g}}_m^T{{\mathbf{T}}_k}} \right)}  + {\zeta_{gm}}\sum\limits_{k = 1}^K {\text{tr} \left( {{{\mathbf{T}}_k}} \right)},  \hfill \\ 
\end{gathered} 
\end{equation}
where $ {\zeta_{gm}}={2{\delta _{gm}}\left\| {{{{\mathbf{\hat g}}}_m}} \right\| + \delta _{gm}^2}  $. We optimize the upper bound of the total interference power, which is obtained as
\begin{equation}
\begin{gathered}
  \text{tr} \left( {{{\mathbf{G}}^T}{\mathbf{\tilde T}}{{\mathbf{G}}^*}} \right)=\sum\limits_{k = 1}^K {{\text{tr}}\left( {{{\mathbf{G}}^*}{{\mathbf{G}}^T}{{\mathbf{T}}_k}} \right)}  = \sum\limits_{m = 1}^M {{{\left| {{\mathbf{g}}_m^T\sum\limits_{k = 1}^K {{{\mathbf{t}}_k}} {d_k}} \right|}^2}}  \hfill \\
   \le\sum\limits_{k = 1}^K {{\text{tr}}\left( {{{{\mathbf{\hat G}}}^*}{{{\mathbf{\hat G}}}^T}{{\mathbf{T}}_k}} \right)}  + \sum\limits_{m = 1}^M {{\zeta _{gm}}} \sum\limits_{k = 1}^K {\text{tr} \left( {{{\mathbf{T}}_k}} \right)}. \hfill \\ 
\end{gathered}
\end{equation}
For the SINR constraint with partially known channel $\bf H$ and $\bf F$, a worst-case approach is considered to guarantee that the solution is robust to all the uncertainties. Based on the triangle inequality, the maximum interference power from radar to the \emph{i}th user is given as
\begin{equation}
\begin{gathered}
    P_R{{\left\| {{{\mathbf{f}}_i}} \right\|}^2} = P_R{{\left\| {{{{\mathbf{\hat f}}}_i} + {{\mathbf{e}}_{fi}}} \right\|}^2} \hfill \\
    \le P_R{{\left( {\left\| {{{{\mathbf{\hat f}}}_i}} \right\| + \left\| {{{\mathbf{e}}_{fi}}} \right\|} \right)}^2} \le P_R{{\left( {\left\| {{{{\mathbf{\hat f}}}_i}} \right\| + {\delta _{fi}}} \right)}^2}.
\end{gathered}
\end{equation}
Following the well known S-procedure\cite{boyd2004convex}, the upper bound minimization with worst-case constraints is given as
\begin{equation}
\begin{gathered}
  {{\cal P}_3}:\mathop {\min }\limits_{{{\mathbf{T}}_i},{s_i}}\sum\limits_{i = 1}^K {{\text{tr}}\left( {{{{\mathbf{\hat G}}}^*}{{{\mathbf{\hat G}}}^T}{{\mathbf{T}}_i}} \right)}  + \sum\limits_{m = 1}^M {{\zeta _{gm}}} \sum\limits_{i = 1}^K {\text{tr} \left( {{{\mathbf{T}}_i}} \right)}  \hfill \\
  s.t.\;\left[ {\begin{array}{*{20}{c}}
  {{\mathbf{\hat h}}_i^T{{\mathbf{Q}}_i}{{{\mathbf{\hat h}}}_i^*} - {\Gamma _i}{\beta _i} - {s_i}\delta _{hi}^2}&{{\mathbf{\hat h}}_i^T{{\mathbf{Q}}_i}} \\ 
  {{{\mathbf{Q}}_i}{{{\mathbf{\hat h}}}_i^*}}&{{{\mathbf{Q}}_i} + {s_i}{\mathbf{I}}} 
\end{array}} \right] \succeq 0, \hfill \\
  \;\;\;\;\;\;\;{{\mathbf{T}}_i} \succeq 0,{{\mathbf{T}}_i} = {\mathbf{T}}_i^*,{\text{rank}}\left( {{{\mathbf{T}}_i}} \right) = 1,{s_i} \ge 0,\forall i, \hfill \\
  \;\;\;\;\;\;\;\sum\limits_{i = 1}^K {{\text{tr}}\left( {{{\mathbf{T}}_i}} \right)}  \le {P_0}, \hfill \\ 
\end{gathered} 
\end{equation}
where $ {{\mathbf{Q}}_i} = {{\mathbf{T}}_i} - {\Gamma _i}\sum\limits_{n = 1,n \ne i}^K {{{\mathbf{T}}_n}}, {{\beta _i}={P_R}{{\left( {\left\| {{{{\mathbf{\hat f}}}_i}} \right\| + {\delta _{fi}}} \right)}^2} + \sigma _C^2} $. Similar to problem $ {\cal{P}}_2 $, by dropping the rank constraint, the non-convex problem $ {\cal{P}}_3 $ becomes a standard SDP and can be solved by SDR.
\subsection{Weighted Minimization for Imperfect CSI}
It is important to note that the upper bound minimization can only guarantee that the obtained beamformer does not generate strong interference, and it may not perform well for all the realizations of the interference channel $\bf G$. Here we use a weighted minimization for the case. Consider that $\bf G$ is perfectly known to the BS, thus the actual power of the interference can be minimized. On the contrary, if BS has no knowledge of $\bf G$, the best strategy is to minimize the transmit power since a large power may cause higher interference. Obviously, the case with the partially known $\bf G$ falls in between these two extreme cases. In other word, BS knows the estimated form of the interference power $\sum\limits_{i = 1}^K {{\text{tr}}\left( {{{{\mathbf{\hat G}}}^*}{{{\mathbf{\hat G}}}^T}{{\mathbf{T}}_i}} \right)}$, and the uncertainty about it, which is decided by the norm bound $\delta_{gm}$. If $\delta_{gm}$ is large, we are more uncertain about the estimated interference, so we put more weight on minimizing the transmit power. Based on this, we rewrite ${\cal P}_3$ as
\begin{equation}
\begin{array}{*{20}{l}}
  {{\cal P}_4}:\mathop {\min }\limits_{{{\mathbf{T}}_i},{s_i}}\sum\limits_{i = 1}^K {{\text{tr}}\left( {{{{\mathbf{\hat G}}}^*}{{{\mathbf{\hat G}}}^T}{{\mathbf{T}}_i}} \right)}  + {{\phi (\delta_{g1},...,\delta_{gm})}} \sum\limits_{i = 1}^K {\text{tr} \left( {{{\mathbf{T}}_i}} \right)}  \hfill \\
  s.t. \;\;\text{The same constraints with } {{\cal P}_3},
\end{array} 
\end{equation}
where $\phi (\delta_{g1},...,\delta_{gm})$ is an increasing function of error bounds. It can be observed that the upper bound minimization is a particular case of the weighted minimization with its weight function equalling to $\sum\limits_{m = 1}^M {{\zeta _{gm}}}$. Our results below show that this weight function is in general large, which means that it puts too much weight on minimizing the BS power while the uncertainty about the estimated channel $\mathbf{\hat G}$ is small.
\section{Numerical Results}
\begin{figure}
    \centering
    \subfloat[]{\includegraphics[width=0.492\columnwidth]{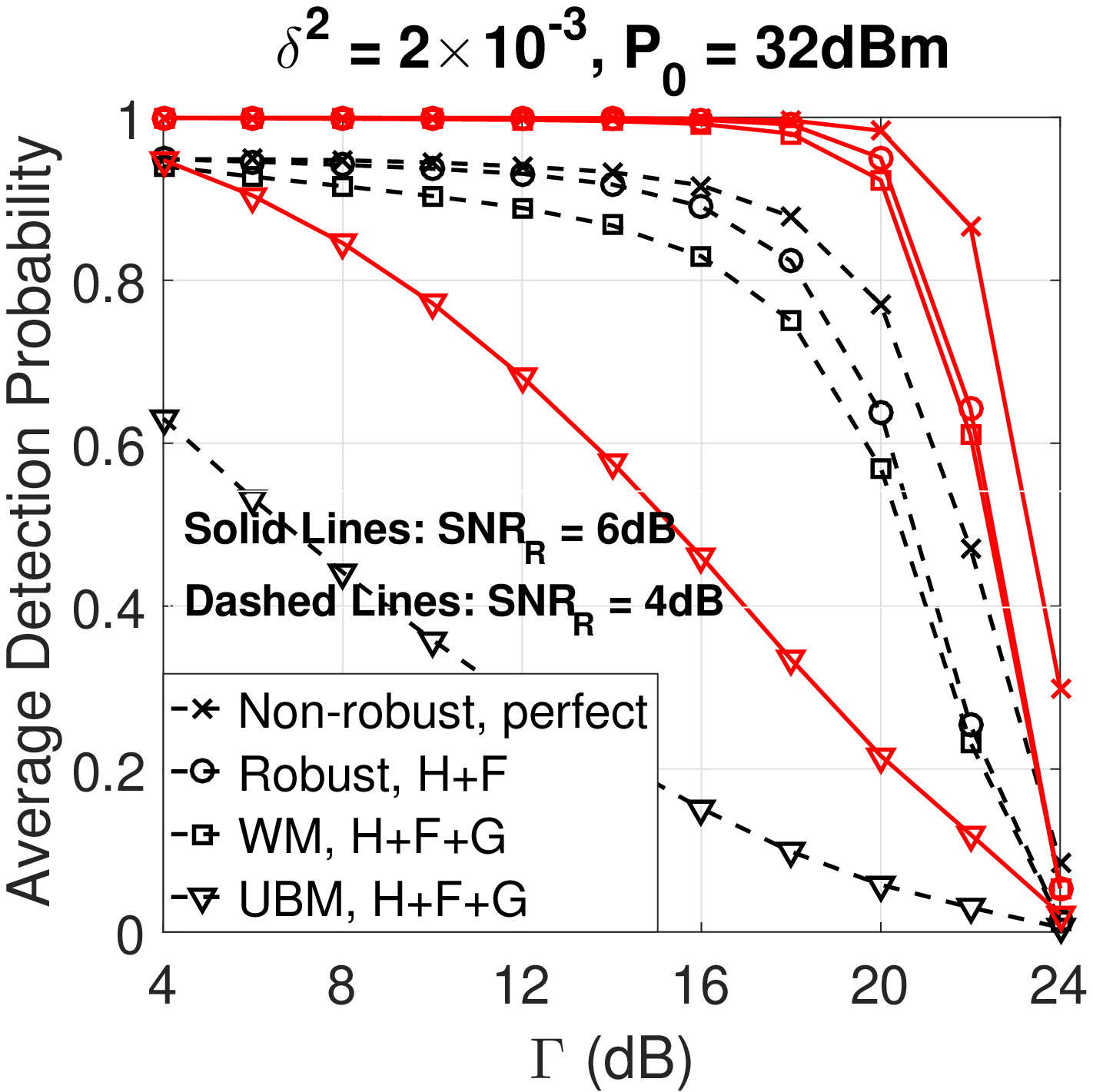}}
    \hspace{.01in}
    \subfloat[]{\includegraphics[width=0.492\columnwidth]{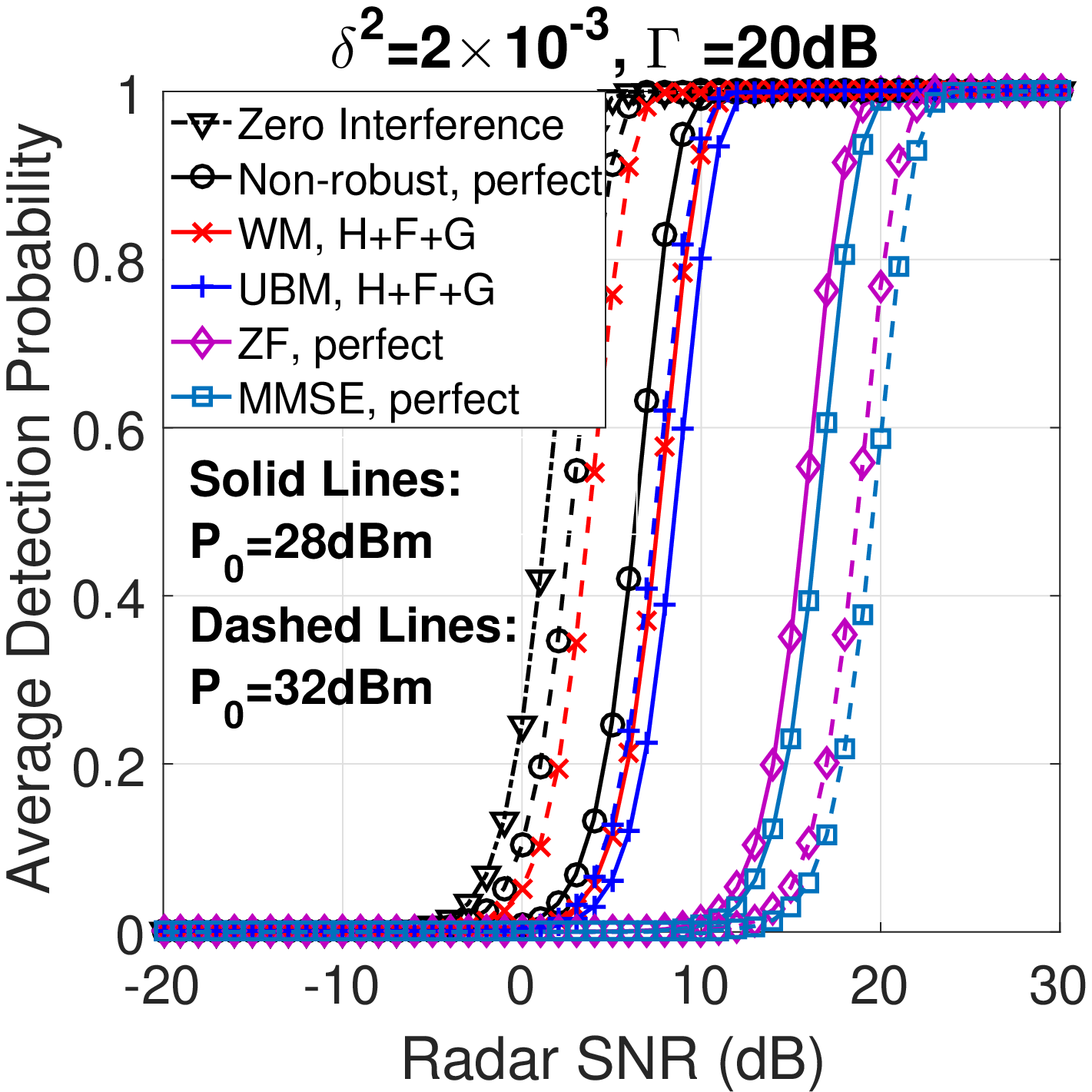}}
    \caption{Numerical results. (a) Average detection probability vs. SINR level for different cases, $\delta^2 = 2\times10^{-3},P_0=32\text{dBm}, P_{FA}=10^{-5}$; (b) Average detection probability vs. radar SNR for different cases, $\delta^2 = 2\times10^{-3}, \Gamma = 20 {\text{dB}}, P_{FA}=10^{-5}$.}
    \label{fig:2}
\end{figure}
In this section, numerical results based on Monte Carlo simulations have been provided to validate the effectiveness of the proposed method. Without loss of generality, each entry of the channel matrices follows the standard complex Gaussian distribution. We assume that radar uses a Uniform Linear Array (ULA) and has unit power. For convenience, we set $ \Gamma_i = \Gamma, \delta_{hi} = \delta_{fi} = \delta_{gm} = \delta, \forall i, m $. While it is plausible that the benefits of the proposed scheme extend to various scenarios, here we assume $ N=8 $, $ K=4 $, $ M=4 $, $\delta^2 = 2\times10^{-3}$ and $P_{FA}=10^{-5}$ in all simulations. The power of all the noise vectors are set as $\sigma_R^2=\sigma_C^2=0\text{dBm}$. The radar SNR is defined as ${\text{SNR}}_R = \frac{L|\alpha|^2P_R}{\sigma_R^2}$\cite{1703855}. And we use a weight function $\phi(\delta)=\frac{\delta}{M}$ in the weighted minimization.
\\\indent In Fig. 2 (a), the average detection probability $ P_D $ with increasing SINR level is shown for $ P_0=32\text{dBm}$. The legends denote the beamformer used and the CSI state assumed, where `Non-robust', `Robust', `UBM' and `WM' denote beamforming optimization ${\cal P}_2$, ${\cal P}_2$ using robust constraints, upper bound minimization ${\cal P}_3$ and weighted minimization ${\cal P}_4$ respectively. `Perfect', `$\bf H+\bf F$' and `$\bf H+\bf F+\bf G$' denote the case of perfect CSI, the case that $\bf H$ and $\bf F$ suffer from CSI errors while $\bf G$ is perfectly known and the case that all channels are with errors. It can be seen that $P_D$ decreases with the growth of $\Gamma$, which formulates the trade-off between the performance of radar and downlink communications. Note that the case `Robust, $\bf H+\bf F$' and the weighted minimization show detection performance close to that for the perfect CSI case. Nevertheless, a significant performance loss occurs for upper bound minimization since it puts too much weight on minimizing the transmit power when $\delta$ is small. Similar results have been shown in Fig. 2 (b), where $P_D$ with increased radar SNR for different power budget has been given with $\Gamma = 20\text{dB}$. The idealistic case that BS causes no interference to radar has been provided as a reference. Once agian, the weighted minimization outperforms the upper bound minimization. To further prove the optimality of the proposed methods, we also show the performance of Zero-forcing (ZF) and Minimum Mean Squared Error (MMSE) beamforming methods with perfect CSI. Unsurprisingly, even the upper bound minimization achieves a far better performance than the conventional methods. It can be also observed that larger $P_0$ leads to higher $P_D$ for the proposed methods due to the extension of the feasible domain. 
\section{Conclusion}
A beamforming approach has been introduced to facilitate the coexistence between downlink MU-MIMO communication system and MIMO radar. Given a target communication link SINR and the transmit power budget of BS, the proposed beamformer optimizes the radar performance. The proposed optimization has also been made robust to CSI errors by the upper bound minimization and the weighted minimization. The trade-off between radar and communication performance as well as the effectiveness of the proposed approach have been revealed by numerical simulations.

\ifCLASSOPTIONcaptionsoff
  \newpage
\fi



\bibliographystyle{IEEEtran}
\bibliography{IEEEabrv,Liu_WCL2017-0209}

\end{document}